\documentclass[12pt]{iopart}
\usepackage{amsfonts}
\begin{document}
\title [ Plasma waves driven by gravitational waves
in an expanding universe]{Plasma waves driven by gravitational waves
in an expanding universe }
\vspace{1cm}
\author{D.B. Papadopoulos}{Present address:
University of Thessaloniki, Department of Astronomy,Thessaloniki Greece 54006, Greece}
\begin{abstract}
In a Friedmann-Robertson-Walker (FRW) cosmological model with
zero spatial curvature, we consider the interaction of the
gravitational waves with the plasma in the presence of a weak
magnetic field. Using the relativistic hydromagnetic equations it
is verified that large amplitude magnetosonic waves are excited,
assuming that both, the gravitational field and the weak magnetic
field do not break the homogeneity and isotropy of the considered
FRW spacetime.
\end{abstract}
\maketitle

\section{Introduction}

In recent years, considerable progress has been made in
understanding plasma phenomena in the radiation era (plasma
epoch), a period of the expanding universe which has particular
interest. It is known that after the end of the inflation, the
universe enters standard Friedman-Robertson-Walker(FRW) stage and
the wavelength of the large-scale perturbations increases less
rapidly than the Hubble length which, at that time corresponds to
the particle horizon (Zimdahl W 1997). It is known that at the
end of the inflationary era density perturbations and
gravitational waves(GW) are produced (Brevik I and Sandvik H B
2000, Lyth D H 1996). The density perturbations are thought to be
responsible for the formation of large-scale structure along with
a possible GW contribution. Moreover, the density perturbations
due to the inflation are responsible for the observed
anisotropies of the Cosmic Microwave Background Radiation(CMBR)
(Kandaswamy Subramanian and Barrow J D 1998, Koh S and Lee C H
2000). On the other hand, it is argued (see Dimopoulos K et al
2001) that at the end of the inflationary epoch large-scale
magnetic fields and inhomogeneities could be generated. The
existence of large-scale magnetic fields in the early universe
may have important consequences on the large scale structure
formation process as shown in (Maartens R et all C 2001, Tsagas G
C 2001, Grasso D and Rubinstein H R 2001, Tsagas G C and Maartens
R 2000). The history of the magnetic fields from magnetogenesis
in the early universe to the present time with the evolution and
damping of the magnetic fields up to recombination era, is
presented in (Jedamzik K et al 2000, Olinto V 1998). All this
knowledge related to the density perturbations and magnetic
inhomogeneities in the early stages of the universe, help us to
construct possible scenarios to discuss plasma phenomena (see
Holcomb K A and Tajima T 1989, Holcomb K A 1990, Dettmann et al
1993, Sahni V 1989, Ford L H and Parker L 1977, Caprini C and
Durrer R 2001, Brodin G,  Marklund M 1999, Marklund M et all 2000,
Brodin G et al 2001, Brodin G  et al 2001). In particular, from
$t=10^{-2}s$ to the time of recombination the ordinary matter
content will be in the form of plasma. From approximately
$t=10^{-2}s$ to $t=1s$, the plasma is dominated by electrons and
positrons while from $t=1s$ to the time of recombination at
roughly $t=10^{13}s$, the plasma consists mostly of electrons and
protons with a admixture of ions of light elements. These plasmas
are believed to be in thermal equilibrium, or at last strongly
coupled, with photons. It is the photons however, which dominates
the energy density of the universe until the time $\approx 10^{2}
s$ Thus most of the period prior to the recombination is
traditionally called the "radiation era" of the universe. In this
paper we will focus on the radiation epoch of the universe and
try to discuss, whether the existence of the GW in a FRW space
time with zero spatial curvature and small magnetic field may
produce instabilities. These kind of problems are of particular
interest since astrophysicists have began to discuss
finite-amplitude wave propagation in hydromagnetic systems in the
presence of GW (Papadopoulos D et al 2001, Papadopoulos D Vlahos
L and Esposito F P 2001). Long ago Papadopolos and Esposito
(Papadopoulos D Esposito F P 1982), derived the exact equations
governing finite amplitude wave propagation in hydromagnetic
media within the general theory of relativity, in the so called
Cowling approximation (Cowling T G 1941). These equations are of
particular interest because when they are linearized, they are
very useful to discuss stability criteria in the presence of
magnetic fields. Recently, (Papadopoulos et al. 2001) have
derived a dispersion relation in an anisotropic cosmological
model (Thorne K S 1967). They verified that in the case that the
parameter $\gamma$, used to parametrize the equation of state,
approaches $1/3$, the anisotropic cosmological model becomes
isotropic, actually a FRW with a scale factor $S(t) =t^{1/2}$,
and the expansion $\theta$ of the universe prevents any
instability. But in ((Dimopoulos K et al 2001) the authors argue
that in FRW cosmological model, if one starts with small
fluctuations, may obtain growth in those instabilities. The
classical theory of GW perturbations on FRW universe has been
worked out by Lifshitz and discussed by several authors (Ya.B
Zel'dovich and Novikov I D 1983 ). In this paper we consider a
FRW cosmological model,with signature $(-,+,+,+)$ witch is
spatial flat and a weak magnetic field, perturbed by a GW with
two polarization modes. As state by Marklund M et al 2000, the
presence of a weak magnetic field propagating on the background
FRW model couple to the gravitational wave producing a pulse of
gravitationally induced electromagnetic wave. In principle, this
breaks the homogeneity and isotropy of the FRW spacetime.
However, it can be shown that a weak magnetic field is consistent
with such a background provided that the spatial variations it
generates are negligible. Therefore, we assume that both, the
weak magnetic field and the GW, do not break the homogeneity and
isotropy of the FRW space time. Applying the formalism presented
in (Ya.B Zel'dovich and Novikov I D 1983, Brandenburg A et al
2000, Weinberg S 1972) in our scenario, we verify that density
fluctuations may be obtained.

\section{The Equations}
The exact equations governing finite-amplitude wave propagation
in hydromagnetic media in the frame of general relativity have
been discussed in (Papadopoulos D Esposito F P 1982, Papadopoulos
D. et al.2001, Papadopoulos D Vlahos L and Esposito F P 2001). For
completeness we recast the relevant equations. We start with the
Einstein field equations
\begin{eqnarray}
R_{ab}-\frac{1}{2}g_{ab}R&=-&\kappa T_{ab},
\end{eqnarray}
with
\begin{eqnarray}
T_{;b}^{ab}&=&0
\end{eqnarray}
For simplicity we obtain $c=1$ and $\kappa=\frac{8\pi G}{c^4}=1$.
Taking the covariant divergence of the Bianchi identities we
obtain
\begin{eqnarray}
(R^{ab}-\frac{1}{2}R g^{ab})_{;ab}&=&0
\end{eqnarray}
where $R_{ab}=R_{abc}^{c}$ is the Ricci tensor and $R_{abcd}$ is the curvature tensor defined
by
\begin{eqnarray}
u_{a;bc}-u_{a;cd}&=&u_{d} R_{abc}^d
\end{eqnarray}
For a unit time-like vector we choose $u_{a}u^{a}=-1$ and our hydromagnetic system will
be specified by the following choice for the energy-momentum tensor
\begin{eqnarray}
T^{ab}&=&(\epsilon+\frac{H^2}{2})u^au^b+(p+\frac{H^2}{2})h^{ab}-H^aH^b
\end{eqnarray}
with
\begin{eqnarray}
h^{ab}&=&g^{ab}+u^au^b,~~\epsilon=\rho+\rho\Pi
\end{eqnarray}
where $u^a$ is the fluid velocity, $\rho$ the mass density,
$\rho\Pi$ the internal energy density, $p$ the pressure of the
fluid and $H^a$ is the prevailing magnetic field as measured by
an observer co-moving with $u^{a}$. Furthermore we introduce the
expansion $\theta=u_{;a}^a$, the shear $\sigma_{ab}=h_a^ch_b^d
u_{(c;d)}-\frac{1}{3}\theta h_{ab}$ and the twist
$\omega_{ab}=h_a^c h_b^d u_{[c;d]}$, where the round bracket
denotes symmetrization while the square bracket
antisymmetrization. From Eq.(1) and Eq.(5) we find
\begin{eqnarray}
R^{ab}&=&-\{\frac{1}{2}(\epsilon+3p+H^2)u^{a}
u^{b}+\frac{1}{2}(\epsilon-p+H^2)h^{ab}-H^{a} H^{b}\}
\end{eqnarray}
Substituting Eq.(7) into Eq.(3) we have
\begin{eqnarray}
\ddot x+g^{ab} \left (p+\frac{H^2}{2}\right)_{;ab} +2\dot{x}\theta+x_{;a}
\dot{u}^a+x(\dot{\theta}+\theta^2+\dot{u}_{;a}^{a})-(H^a
H^b)_{;ab}&=&0
\end{eqnarray}
where $\dot{u}^a=u_{;c}^a u^c$ and $x=\epsilon+p+H^2$ (Notice,that
Eq.(8) can be obtained directly from Eq.(2) and Eq.(5)).
Raychaudhuri's equation in the form
\begin{eqnarray}
\dot{u}_{;a}^a&=&\dot{\theta}+\frac{\theta^2}{3}+2(\sigma^2-\omega^2)+\frac{1}{2}(\epsilon+3p+H^2)
\end{eqnarray}
where $2\sigma^2=\sigma^{ab}\sigma_{ab}$,
$2\omega^2=\omega^{ab}\omega_{ab}$ and dot means covariant
derivative along $u^{a}$ allows us to write the Eq.(8) as
\begin{eqnarray}
\ddot x +g^{ab} \left(p+\frac{H^2}{2} \right)_{;ab} +2\dot{x}\theta+2x\dot{\theta}+x_{;a} \dot{u}^a\nonumber\\
+2x(\frac{2\theta^2}{3}+\sigma^2-\omega^2)+\frac{1}{2}x(\epsilon+3p+H^2)-(H^a H^b)_{;ab}&=&0
\end{eqnarray}
In order to make further progress with Eq.(10) we must use the equations of motion
for the fluid and Maxwell's equations. These equations are
\begin{eqnarray}
0=T_{;b}^{ab}=\dot{x} u^a+x\dot{u}^a+x\theta
u^a+(p+\frac{H^2}{2})_{;b}g^{ab}-(H^a H^b)_{;b}
\end{eqnarray}
and
\begin{eqnarray}
\dot{H}^a&=&(\sigma_b^a+\omega_b^a-\frac{2}{3}\delta_b^a\theta)H^b+\frac{1}{\epsilon+p}p_{;b}H^b u^a
\end{eqnarray}
The last equation (12) may be written in the following form
\begin{eqnarray}
\dot{\frac{\mu H^2}{8\pi}}&=&\frac{\mu}{4\pi}\sigma_{ij} H^i
H^j-\frac{4\theta}{3}(\frac{\mu H^2}{8\pi})
\end{eqnarray}
where $\mu$ is the permeability.
Equations (11) and (12) imply that
\begin{eqnarray}
\ddot{\epsilon}+(\dot{\epsilon}+\dot{p})\theta+(\epsilon+p)\dot{\theta}&=&0
\end{eqnarray}
Recalling that $x=\epsilon+p+H^2$ we find that Eq.(10) takes the form
\begin{eqnarray}
(\epsilon-\frac{H^2}{2})_{;ab} u^a
u^b=h^{ab}(p+\frac{H^2}{2})_{;ab}+2\dot{(H^2\Theta)}-(H^aH^b)_{;ab}\nonumber\\
+2x(\frac{2\theta^2}{3}+\sigma^2-\omega^2-\dot{u}^a\dot{u}_{a})+\frac{x}{2}(\epsilon+3p+H^2)\nonumber\\
+2\dot{u}_{a}(H^aH^b)_{;b}+(H^2)_{;a}\dot{u}^a \end{eqnarray}
The Eqs. (11),(12) and (15) may be applied to the investigation of
perturbation effects and hence to the study of the linearizied
stability criteria.

\section{The Perturbed Equations}

Following the well established method by (Hawking 1966), used
also in (Papadopoulos D Esposito F P 1982, Papadopoulos D et al.
2001, Papadopoulos D Vlahos L and Esposito F P 2001), we consider
the perturbations as follows:

\begin{eqnarray}
\epsilon&=\epsilon_0+\delta \epsilon,& p=p_0+\delta p
\end{eqnarray}
\begin{eqnarray}
u^{\mu}&=u_{0}^{\mu}+\delta u^{\mu},& H^{\mu}=H_{0}^{\mu}+\delta
H^{\mu}
\end{eqnarray}
where the index $0$ means unperturbed quantities.

Taking into account the Eqs.(16),(17), we perturbed the equations
Eqs.(15),(11) and (12) respectively, keeping only linear terms in
the perturbed equations and assuming that $\delta g_{ab}\neq 0$.
Thus we have
\begin{eqnarray} & &\delta[(\epsilon-\frac{H^2}{2})_{;ab}]u^a u^b-\delta \Gamma_{ab}^k
(\epsilon-\frac{H^2}{2})_{,k} u^a
u^b\nonumber\\&+&(\epsilon-\frac{H^2}{2})_{;ab}
(u^a\delta u^b+u^b\delta u^a)+(\delta\epsilon-H^c\delta H_{c})_{;ab}u^a u^b\nonumber\\
&=&\delta g^{ab} p_{,ab}^{*}+(u^a\delta u^b+u^b \delta
u^a)p_{ab}^{*}+
h^{ab}(\delta p_{,ab}^{*}-\delta \Gamma_{ab}^k p_{,k}^{*})\nonumber\\
&+&2\delta [(H^2)_{,c}
\Theta+H^2\Theta_{,c}]u^c-H^b[2\delta(H_{,ab}^a)+
H^k\delta(\Gamma_{ak,b}^a)+2\delta \Gamma_{ak}^a H_{,b}^k]\nonumber\\
&-&2H_{,ab}^a\delta H^b-2\delta \Gamma_{al}^a H^l H_{,b}^b-
\delta (H_{,b}^a) H_{,a}^b-H_{,b}^a\delta (H_{,a}^b\nonumber\\
&-&\delta \Gamma_{bl}^a H^l H_{,a}^b-H^aH^c\delta \Gamma_{ac,b}^b-
\delta \Gamma_{ac}^b H_{,b}^c H^a+2x [\sigma \delta \sigma -
u^c u^d(\delta u_{,c}^a u_{a,d}+u_{,c}^a\delta u_{a,d}\nonumber\\
&+&\delta \Gamma_{cl}^a u^l u_{a,d}-u_{,c}^a u_m\delta
\Gamma_{ad}^m)-
(u^d\delta u^c+u^c\delta u^d)u_{,c}^a u_{a,d}]\nonumber\\
&+&\frac{1}{2}(\epsilon+3p+H^2)(\delta \epsilon+\delta p+
\delta H^2)+\frac{1}{2}x(\delta \epsilon+3\delta p+ \delta H^2)\nonumber\\
&+&[2\delta u^c u_{a,c}+2u^c(\delta u_{a,c}- \delta \Gamma_{ac}^m
u_m) ](H^a\delta H_b+H^b\delta H^a)+
(H^2)_{;a}(\delta u^a)_{;c}u^c\nonumber\\
&+&(H^2)_{,a}u_{,c}^a\delta u^c+(H^2)_{,a} u^c(\delta
u_{,c}^a+\delta \Gamma_{cm}^a u_m)
\end{eqnarray}

The perturbed Maxwell's equations are

\begin{eqnarray} \delta H_{,0}^b+\delta \Gamma_{0l}^b H^l&=&-\delta
\sigma_{a}^b H^a+\frac{2}{3}H^b\delta
\theta-\frac{u^b}{\epsilon+p}\delta p_{,c} H^c \end{eqnarray}

and the perturbed Eqs.(11) give

\begin{eqnarray}
-(\epsilon+p+H^2)\delta u_{,0}^c&=&h_a^c [H^b\delta
H_{,b}^a+\delta
\Gamma_{bl}^a H^b H^l\nonumber\\
&+&H^a\delta H_{,b}^b+\delta \Gamma_{bm}^b H^m H^a]-\delta h^{cb}
p_{,b}^{*}- h^{cb}\delta p_{,b}^{*}
\end{eqnarray}

We will apply these equations to a spacial flat FRW space-time
given by the equation (see Ya.B Zel'dovich and Novikov I D 1983)
\begin{eqnarray}
ds^2=-c^2 dt^2+S^2[(1+\frac{h_1}{S^2}) dx^2+(1-\frac{h_1}{S^2})
d^2 y+ d^2 z]+2h_2dxdy
\end{eqnarray}
where $S=S(t)$ is the scale factor of the FRW universe,
parametrazing the universe expansion with
$h_1=h_{+}(t,z),h_2=h_{\times}(t,z)$ are the two components of
the GW corresponding to the two possible polarizations. For the
metric (21) the non-zero $\Gamma_{ab}^{c}$ are
\begin{eqnarray}
\Gamma_{11}^{0}= S S_{,0}+\frac{1}{2} h_{1.0},  \Gamma_{12}^{0}=\frac{1}{2}h_{2,0},\nonumber\\
\Gamma_{22}^{0}=S S_{,0}-\frac{1}{2} A_{1,0},  \Gamma_{33}^{0}= S S_{,0},\nonumber\\
\Gamma_{01}^{1}= \frac{ S_{,0}}{S}+\frac{1}{2}(\frac{h_{1,0}}{S^2}-2\frac{h_1 S_{,0}}{S^3}),\nonumber\\
\Gamma_{02}^{1}=+\frac{1}{2}(\frac{h_{2,0}}{S^2}-2\frac{h_2 S_{,0}}{S^3}),\nonumber\\
\Gamma_{13}^{1}=\frac{1}{2 S^2} h_{1,z0},  \Gamma_{23}^{1}= \frac{1}{2 S^2} h_{2,z0}, \Gamma_{01}^{2}=+\frac{1}{2}(\frac{h_{2,0}}{S^2}-2\frac{h_2 S_{,0}}{S^3}),\nonumber\\
\Gamma_{02}^{2}=\frac{ S_{,0}}{S}-\frac{1}{2}(\frac{h_{1,0}}{S^2}-2\frac{h_1 S_{,0}}{S^3}), \Gamma_{13}^{2}= \frac{1}{2} \frac{h_{2,z}}{S^2},  \Gamma_{23}^{2} =-\frac{1}{2 S^2} h_{1,z},\nonumber\\
\Gamma_{03}^{3}= \frac{S_{,0}}{S}, \Gamma_{11}^{3}= -\frac{1}{2 S^2} h_{1,z} , \Gamma_{12}^{3}= -\frac{1}{2 S^2} h_{2,z} , \Gamma_{22}^{3} =\frac{1}{2 S^2} h_{1,z}
\end{eqnarray}
In an expanding universe described by the metric (21), we obtain
$S(t)=t^{1/2}$ and consider an equation of state
$p=\frac{1}{3}\epsilon$(radiation era). We assume that
$u^{a}=(-1,0,0,0)$, $H^{a}=\frac{1}{S^3}(0,H_0^1,H_0^2,H_0^3)$
(Holcomb  1990), with
$H_0^{a}=const$,$H_0^2=(H_0^1)^2+(H_0^2)^2+(H_0^3)^2$, $\delta
H^{a}=\frac{1}{S^2}(0,\delta H^1, \delta H^2, \delta H^3)$,
$\delta u^{a}=(\delta u^0,\delta u^1,\delta u^2,\delta u^3)$ and
$k^{a}=(0,0,0,k)$ where $k$ is the magnitude of the comoving
wavelength (momentum). Thus, the perturbed Maxwell's equations
(19) take the form
\begin{eqnarray}
\delta H_{,0}^1=\theta \delta H^1+H^1\delta \theta-\frac{1}{3}(\theta H^1-H_{,0}^1))\delta u^0 \nonumber\\
-\frac{1}{2} H^3\delta u_{,3}^1-\frac{\theta}{6S^2} (H^1 h_1),\nonumber\\
\delta H_{,0}^2=\theta \delta H^2-\frac{1}{2}H^3\delta u_{,3}^2-\frac{\theta}{6S^2}(H^1 h_2),\nonumber\\
\delta H_{,0}^3=\theta \delta H^3+H^3\delta \theta-\frac{1}{3}\theta H^3\delta u^0-\frac{1}{2} H^1\delta u_{,3}^1
-H^3\delta u_{,3}^3
\end{eqnarray}
and the equations of motion (20), reduce to the equations
\begin{eqnarray}
\delta u_{,0}^0=0\nonumber\\
x\delta u_{,0}^1=\frac{2}{3}x\theta \delta u^1-H^3\delta H_{,3}^1-H^1\delta H_{,3}^3-\frac{(H^1 H^3)}{S^2} h_{1,3},\nonumber\\
x\delta u_{,0}^2=\frac{2}{3}x\theta \delta u^2-H^3\delta H_{,3}^2-\frac{(H^1 H^3)}{S^2} h_{2,3},\nonumber\\
x\delta u_{,0}^3=\frac{2}{3}x\theta \delta u^3-2H^3\delta H_{,3}^3+\frac{1}{S^2}H_{0\mu}\delta H_{,3}^{\mu}+\delta p_{,3}+(H^1)^2 h_{1,3}
\end{eqnarray}
In this case the Eq.(18) becomes:
\begin{eqnarray}
\delta \epsilon_{,00}-\frac{1}{S^2}\delta p_{,33}-\frac{1}{S^3}H_{0\mu}\delta H_{,00}^{\mu}-\frac{1}{S^5}H_{0\mu}\delta H_{,33}^{\mu}+\frac{2H_0^3}{S^3}\delta H_{,33}^3\nonumber\\
-\theta\delta p_{,0}+\frac{\theta}{3}\frac{H_{0\mu}\delta H_{,0}^{\mu}}{S^3}-[\frac{5\theta^2}{18}+2(\epsilon+2p+\frac{H_0^2}{S^4})]\frac{(H_{0\mu}\delta H^{\mu}}{S^3})\nonumber\\
-4\theta \delta \theta(\epsilon+p-\frac{H_0^2}{3S^2})+\frac{H_0^2}{S^4} \delta \theta_{,0}- \frac{4\theta^2}{3}(\delta \epsilon+\delta p)\nonumber\\
-\frac{1}{2}(\delta \epsilon+\delta p)(\epsilon+3p)-\frac{1}{2}(\delta \epsilon+3\delta p)(\epsilon+p)-\frac{H_0^2}{S^4}(\delta \epsilon+3\delta p)\nonumber\\
= (H^1)^2 h_1 [-\frac{k_{g}^2}{S^2}+\epsilon+2p+\frac{H_0^2}{S^4}-\frac{\theta^2}{3}]
\end{eqnarray}
It is known that in an expanding universe the perturbations can
not be solutions for the GW, because the amplitude of such a wave
must decay in time and its frequency must redshift. Thus, writing
the wave equation for a free wave in a cosmological background
and requiring a propagating solution, the spatial part has the
usual form $\exp{(ikz)}$, while the time depended part has the
form $\exp{(i\omega t/S)}$.  In the plasma period, we assume that
the perturbations, Eqs.(16),(17), have the form (Ya.B Zel'dovich
and Novikov I D 1983):  $\delta
H^{a}=\frac{1}{S^2}\exp{(ikz-i\omega t)}$, $\delta
\epsilon=\frac{1}{S^4}\exp{(ikz-i\omega t)}$, $\delta
u^{a}=\delta u_{0}^{a}\exp{(ikz-i\omega t)}$ and  the two
components of the GW (Brandenburg A et al 2000, Weinberg S 1972)
$h_1,h_2=\frac{1}{S}\exp{(ik_gz-i\omega_g t)}$ where
$\omega=\frac{\omega_i}{S}$, $\omega_g=\frac{\omega_{gi}}{S}$ and
$\omega_i$, $\omega_{gi}$ are the frequencies at the initial time
$t_i<t$. Under these considerations, Eq.(25) yields a general
dispersion relation for the coupling of GW with MHD waves which
is:
\begin{eqnarray}
\Lambda R_1\delta \epsilon+\frac{1}{S^3} H_{0 a}\delta H^{a}
R_2=h_1 \Lambda (H^1)^2 R_3
\end{eqnarray}
where
\begin{eqnarray}
\Lambda=\frac{i\omega}{2}(1-\frac{2u_A^2}{S^4})+\frac{\theta}{3}(1+\frac{4u_A^2}{S^4})\nonumber\\
R_1=-\frac{\omega^2}{4}(1+\frac{u_A^2}{S^4})+\frac{i\omega}{2}\theta(1+5c_s^2-10\frac{u_A^2}{S^4})+\frac{k^2c_s^2}{S^2}\nonumber\\
-\frac{1}{2}(\epsilon+3p)(1+c_s^2)-\frac{1}{2}(\epsilon+p)(1+3c_s^2)-\frac{H_0^2}{S^4}(1+3c_s^2)\nonumber\\
R_2=\frac{\omega^2}{4}(1-\frac{u_A^2}{S^4})-\frac{8i\omega u_A^2}{2S^4}+\frac{k^2}{S^2}-2(\epsilon+2p+\frac{H_0^2}{S^4})
+\theta^2(\frac{28}{9}\frac{u_A^2}{S^4}-\frac{47}{3})\nonumber\\
R_3=-\frac{k_g^2}{S^2}+\epsilon+2p+\frac{H_0^2}{S^4}-\frac{5i\theta \omega_g}{2}+8\theta^2\nonumber\\
-\frac{u_A^2}{S^4}[\frac{\omega_g^2}{4}-\frac{9i\theta \omega_g}{6S}+\frac{16}{3}\theta^2]
\end{eqnarray}
and $u_A^2=\frac {\upsilon_a^2}{1+\upsilon_A^2} $,
$\upsilon_A^2=\frac{H^2}{\epsilon}$ is the Alfven speed and $c_s$
is the sound speed. It is important to emphasize that, the second
component of GW does not appear in Eq.(26). Besides, the right
hand side of Eq.(26) is proportional of $H^1$. This suggests that
the dispersion relation (26) may be used for waves that
correspond to the magnetosonic mode ($H^1\neq 0, H^3=0).$

\section{ The dispersion relation}
Furthermore, obtaining $H^1\neq 0$ and $H^3=0$, the Maxwell's equations and the equations of motion give the
following results respectively: $\delta H^3=\delta H^2=0$,
\begin{eqnarray}
\Lambda H_1\delta H^1=-\frac{u_A^2}{S^4}(\frac{\theta}{3}-\frac{i\omega}{2})(1+c_s^2)\delta \epsilon+h_1\frac{H_0^2}{S^6}[-\frac{u_A^2}{S^4}(\frac{4\theta}{3}-\frac{i\omega_g}{2})+\frac{\theta}{3}]
\end{eqnarray}
and $\delta u^1=0$,$\delta u^2=0$. We notice that in all
equations above equations we have taken $c=1$, $\kappa=8\pi
G/c^4=1$ and we have put $H=B \to \sqrt{4\pi}H$ (from Gaussian
units), where $\kappa$ is the coupling constant in Einstein's
field equations. Substituting Eq.(28) into Eq.(26) and using
ordinary units, we obtain
\begin{eqnarray}
(A_r+iA_i)\delta \epsilon=h_1\frac{(H^1)^2}{4\pi }(B_r+i B_i)
\end{eqnarray}
where
\begin{eqnarray}
A_{r}=-(\frac{\omega}{2})^2\Xi_{2}+\Xi_0, A_{i}=-(\frac{\omega}{2})^3 \Xi_3+(\frac{\omega}{2})\Xi_{1},\nonumber\\
B_{r}=-(\frac{\omega}{2})^2 \Pi_{2}+\frac{\omega\omega_g}{4}\Pi_{1}+\Pi_{0},\nonumber\\
B_{i}=-(\frac{\omega}{2})^2 M_2+(\frac{\omega}{2})n M_1+M_0
\end{eqnarray}
with
\begin{eqnarray}
\Xi_{3}=(1+\frac{c^2u_A^2}{S^4})(1-\frac{2u_A^2}{c^2S^4})+\frac{u_A^2}{c^2S^4}(1+\frac{c_s^2}{c^2})(1-\frac{u_A^2}{c^2S^4})\nonumber\\
\Xi_{2}=\theta[(1-\frac{2u_A^2}{c^2S^4})(1+5\frac{c_s^2}{c^2}-10\frac{u_A^2}{c^2S^4})-\frac{1}{3}(1-25\frac{u_A^2}{c^2S^4}(1+\frac{c_s^2}{c^2})]\nonumber\\
\Xi_{1}=(1-\frac{2u_A^2}{c^2S^4})[\frac{k^2c_s^2}{S^2}-\frac{8\pi G}{c^4}\frac{1}{2}(\epsilon+3p)(1+\frac{c_s^2}{c^2})\nonumber\\
-\frac{4\pi G}{c^4}(\epsilon+p)(1+3\frac{c_s^2}{c^2})-\frac{2 G}{c^4}\frac{H_0^2}{S^4}(1+3\frac{c_s^2}{c^2})]\nonumber\\
+\frac{u_A^2}{c^2S^4}(1+\frac{c_s^2}{c^2})[\frac{k^2c^2}{S^2}
-\frac{16\theta\pi G}{3c^4}(\epsilon+2p+\frac{H_0^2}{4\pi S^4})(1+\frac{c_s^2}{c^2})(1-25\frac{u_A^2}{c^2S^4})]\nonumber\\
\Xi_{0}=\frac{\theta}{3}\{(1+\frac{4u_A^2}{c^2S^4})[\frac{k^2c_s^2}{S^2}-\frac{4\pi G}{c^4}(\epsilon+3p)(1+\frac{c_s^2}{c^2})\nonumber\\
-\frac{4\pi G}{c^4}(\epsilon+p)(1+3\frac{c_s^2}{c^2})-\frac{2 G}{c^4}\frac{H_0^2}{S^4}(1+3\frac{c_s^2}{c^2})]\nonumber\\
-\frac{u_A^2}{c^2S^4}(1+\frac{c_s^2}{c^2})[\frac{k^2c^2}{S^2}
-\frac{16\pi G}{c^4}(\epsilon+2p+\frac{H_0^2}{4\pi S^4})+\theta^2(\frac{28u_A^2}{9c^2S^4}-\frac{47}{3})]\}
\end{eqnarray}
\begin{eqnarray}
\Pi_2=\frac{\theta}{3}(1-\frac{u_A^2}{c^2S^4})(1-\frac{4u_A^2}{c^2 S^4})\nonumber\\
\Pi_{1}=5\theta [(1-\frac{2u_A^2}{c^2S^4})(1-\frac{3}{5}\frac{u_A^2}{c^2S^4})-\frac{8}{5}\frac{u_A^2}{c^2S^4}]\nonumber\\
\Pi_{0}=\frac{\theta}{3}\{(1+\frac{4u_A^2}{c^2S^4})[-\frac{k_g^2 c^2}{S^2}-\frac{u_A^2 \omega_g^2}{4c^2S^4}+\frac{8\pi G}{c^4}(\epsilon+2p+\frac{H_0^2}{4\pi S^4})\nonumber\\
+8\theta^2(1-\frac{2u_A^2}{3c^2S^4})]+(1-\frac{4u_A^2}{c^2S^4})[\frac{k^2c^2}{S^2}\nonumber\\
-\frac{16\pi G}{c^4}(\epsilon+2p+\frac{H_0^2}{4\pi S^4})+\theta^2(\frac{28u_A^2}{9c^2S^4}-\frac{47}{3})]\}
\end{eqnarray}
and
\begin{eqnarray}
M_2=\frac{u_A^2\omega_g}{2c^2S^4}(1-\frac{u_A^2}{c^2S^4})\nonumber\\
M_1=(1-\frac{2u_A^2}{c^2S^4}[-\frac{k_g^2c^2}{S^2}-\frac{u_A^2 \omega_g^2}{4c^2S^4}+\frac{8\pi G}{c^4}(\epsilon+2p+\frac{H_0^2}{4\pi S^4})\nonumber\\
+8\theta^2\frac{u_A^2}{c^2S^4}(1-\frac{4u_A^2}{c^2S^4})]\nonumber\\
M_0=-\frac{5\theta^2}{3}\frac{\omega_{g}}{2}(1-\frac{4u_{A}^2}{c^2S^4})(1-\frac{3u_A^2}{5c^2S^4})\nonumber\\
+\frac{u_A^2 \omega_g}{2c^2S^4}[\frac{k^2c^2}{S^2}-2\frac{8\pi G}{c^4}(\epsilon+2p+\frac{H_0^2}{4\pi S^4})+\theta^2(\frac{28u_A^2}{9c^2S^4}-\frac{47}{3})]
\end{eqnarray}
From now on, we will use the notation $n=\frac{\omega}{2}$ and
$n_g=\frac{\omega_g}{2}$.

It is evident that this section 4, from the beginning, relies
heavily on equations and calculations, therefore we shall try to
present our case by splitting this section in two subsections,
one dealing with the gravitational-wave-free case and another one
addressing the impact of the gravitational wave. Notice that the
coefficient of the $\delta \epsilon$ on the left hand side of
Eq.(29) describes the dispersion relation of the fluid. While,
the coefficient of $h_1$ on the right hand side of Eq.(29)
describes the dispersion relation of the gravitational wave.

{\bf 4a. The gravitational-wave-free case}

In the absence of the gravitational wave ($h_{1}=0$), the
dispersion relation (Eq.(29)) is:
\begin{eqnarray}
A_{r}+i A_{i}=0\Rightarrow (-n^2\Xi_2+\Xi_{0})+i(-n^3 \Xi_3+n \Xi_1)=0
\end{eqnarray}
Eq.(34) has three roots, two complex conjugates and one real; we examine one of them,
say the $n_1=n_r+in_i$, with  $n_i<<n_r$, since
its complex conjugate will give the same information regarding the fluid and the real one is not
important from the physical point of view. We substitute $n_1$ in Eq.(34) and the resulting equation
after neglecting non-linear terms of $n_{i}$ is:
\begin{eqnarray}
-(n_r^2+2i n_i n_r)\Xi_2+\Xi_0-i(n_r^3+3in_in_r^2)\Xi_3+i(n_r+in_i)\Xi_1\simeq 0
\end{eqnarray}
From the imaginary part of Eq.(35), we obtain $n_r^2 \simeq -2 n_i\frac{\Xi_2}{\Xi_3}+\frac{\Xi_1}{\Xi_3}$, and $n_i \simeq \frac{\Xi_1\Xi_2-\Xi_0\Xi_3}{2(\Xi_2^2+\Xi_3\Xi_1)}$.
After some elementary calculations, we verify that the condition $n_i << n_r$ is satisfied.
Thus, from Eq.(29) in the absence of the GW, we obtain  Eq.(34) and find the root $n_1$
which has a real part $n_r$, given by the equation
\begin{eqnarray}
n_r^2 \simeq \frac{\frac{10\theta^2}{81}[\frac{17k^2c_s^2}{S^2}+\frac{2}{3}\frac{k^2u_A^2}{S^6}]}{ [\frac{k^2c_s^2}{S^2}+\frac{2}{3}\frac{k^2u_A^2}{S^6}+\frac{400\theta^2}{81}]}+\frac{k^2 c_s^2}{S^2}+\frac{2k^2 u_A^2}{3S^6}
\end{eqnarray}
and an imaginary part, given by the equation
\begin{eqnarray}
n_i \simeq \frac{\frac{\theta}{18}[\frac{17k^2c_s^2}{S^2}+\frac{2}{3}\frac{k^2u_A^2}{S^6}]}{[\frac{k^2c_s^2}{S^2}+\frac{2}{3}\frac{k^2u_A^2}{S^6}+\frac{400\theta^2}{81}]}
\end{eqnarray}
Note, that the two frequencies, $n_r$ and $n_i$, depend on the expansion $\theta$. If, somehow,
the universe would stop to expand, then the real frequency $n_r$ is just the so called magneto-sound
frequency (Jedamzik K et all 2000,  Olinto V 1998),
while the frequency $n_i$ becomes zero. In an expanding universe, an imaginary part of the frequency $n_i$ appear, and the real
frequency $n_r$, is shifted by a term proportional to $\theta^2$.
The existence of the frequency $n_r$ means that before the GW starts to interact with the plasma,
the plasma were oscillating with this $n_r$ normal frequency,
while the frequency $n_i$ corresponds to just a negligible noise.

{\bf 4b. The dispersion relation in the presence of the
gravitational wave.}

Now we consider the case where the GW interacts with plasma. We
assume that the GW is weak and does not affect the background
metric and the frequencies $n_r$ and $n_i$. In this scenario, we
write: $\delta \epsilon \simeq \frac{1}{S^4}\delta \epsilon_0
\exp{i(nt-kz)}$, and $h_1\simeq\frac{1}{S} h_{10} \exp{i(n_g
t-k_gz)}$ (Brandenburg A et all 2000, Weinberg S 1972), where
$\delta \epsilon_0=constant$ and $h_{10}=constant$. Thus, Eq.(29)
is written as
\begin{eqnarray}
\delta \epsilon_{0}\simeq h_{10}e^{i[(n-n_g)t-(k_{g}-k)z]}\frac{H_0^2}{S^3}[\frac{B_r+i B_i}{A_r+i A_i}]
\end{eqnarray}
The term  $T=\frac{B_r+i B_i}{A_r+i A_i}$ takes its maximum value
close to the frequency $n=n_1$. Let $n_g$  and  $k_g$, the frequency and the wave number
of the GW respectively. Close to the frequency $n_1$, we assume that the frequency of the driving GW
$n_g$ coincides with the frequency $n_r=Re(n_1)$. The produced magnetosotic wave has $n_r=n_g=k_gc$ and
the two wave numbers differ by an amount $\Delta k$,  e.g.  $k=k_g+\Delta k$.
Taking into account all the above considerations, Eq.(38) reads
\begin{eqnarray}
\delta \epsilon_0 \approx h_{10}e^{-iz\Delta k}e^{n_i t}\frac{H_0^2}{4\pi S^3} [\frac{B_g}{\Delta k A_k}+\frac{B_k}{A_k}]
\end{eqnarray}
In the linear theory of perturbations we have $\delta \epsilon < \epsilon_0$. Thus, Eq.(39) gives
\begin{eqnarray}
h_{10}e^{-iz\Delta k}e^{n_i t}\frac{H_0^2}{4\pi S^3} [\frac{B_g}{\Delta k A_k}]<\epsilon_0=const.,~ as~~ \Delta k\rightarrow 0 .
\end{eqnarray}
where
\begin{eqnarray}
& &B_g=\frac{\theta}{3}(14 n_g^2-8\frac{k_g^2 u_A^2}{S^6})-n_i(-\frac{k_g^2 c^2}{S^2}+\frac{k_g^2 u_A^2}{S^6})\nonumber\\
&+&i[\frac{13\theta}{3}n_i n_g+n_g(-\frac{k_g^2 c^2}{S^2}+\frac{3k_g^2 u_A^2}{S^6}-\frac{5\theta^2}{3})]
\end{eqnarray}
\begin{eqnarray}
B_k=\frac{\theta}{3}[\frac{2k_g c^2}{S^2}-\frac{4k_g u_A^2}{S^6}]+in_g(\frac{2k_g u_A^2}{S^6})
\end{eqnarray}
and
\begin{eqnarray}
A_k=\frac{\theta}{3}(\frac{2k_gc_s^2}{S^2})-n_i(\frac{2k_gc_s^2}{S^2}+\frac{4 k_g u_A^2}{3S^6})+in_g(\frac{2k_gc_s^2}{S^2}+\frac{4 k_g u_A^2}{3S^6})
\end{eqnarray}
Because of the form of $n_i$ (Eq.(37)) the quantity $e^{tn_i}$ approaches to a constant value, as t
approaches to very large values. Thus, to the approximations we made, the disturbance $\delta \epsilon_0$ does not
exhibit growth. However, the term $e^{tn_i}$ enhanced the amplitude of the oscillation and this
may cause further enhancement on a non-linear approach which we will not discuss in this work.
(The result is in agreement with the result obtained in (Holcomb K A and Tajima T 1989,
Holcomb K A 1990).
Alfven or other MHD oscillations are not enhanced, but magnetosonic frequencies through the interaction with the GW become dominant.
Therefore, it is interesting to know the maximum length which the GW will travel in the plasma
and the maximum time the GW will interact with the plasma.
From  Eq.(40) we have
\begin{eqnarray}
h_{10}e^{n_i t}\frac{H_0^2}{4\pi S^3}[\frac{Re(B_g A_k^{*})}{A_k A_k^{*} \epsilon_0 }] < \Delta k
\end{eqnarray}
Eq.(44) can give us approximately the  $\Delta k$ and subsequently, we can calculate the
maximum distance $L$ over which the two waves may interact coherently. That maximum
$L$ can be calculated in terms of the wavelength $\lambda_g$ through the relation $L=\lambda_{g}(1-\frac{\Delta k}{k_g})$ e.g
\begin{eqnarray}
L<\lambda_g \{1-h_{10}e^{n_i t}\frac{H_0^2}{k_g \epsilon_0}[\frac{I_1\frac{\theta}{3}+n_g^2 I_2}{I_3\frac{2\theta k_g c_s^2}{3S^2}+n_g^2 I_4}]\}
\end{eqnarray}
where
\begin{eqnarray}
& &I_1=\frac{2\theta k_g c_s^2}{3S^2}(14n_g^2-\frac{8k_g^2 u_A^2}{S^6})-n_i[\frac{2k_g c_s^2}{S^2}(-\frac{k_g^2 c^2}{S^2}+\frac{k_g^2 u_A^2}{S^6})\nonumber\\
&+&(14n_g^2-\frac{8k_g^2 u_A^2}{S^6})(\frac{2k_g c_s^2}{S^2}+\frac{4k_g u_A^2}{3S^6})]
\end{eqnarray}
\begin{eqnarray}
I_2=(\frac{2k_g c_s^2}{S^2}+\frac{4k_g u_A^2}{3S^6})(\frac{13\theta n_i}{3}-\frac{k_g^2 c^2}{S^2}+\frac{3k_g^2 u_A^2}{S^6}-\frac{5\theta^2}{3})
\end{eqnarray}
\begin{eqnarray}
I_3=\frac{\theta}{3}\frac{2k_g c_s^2}{S^2}-2n_i(\frac{2k_g c_s^2}{S^2}+\frac{4k_g u_A^2}{3S^6})
\end{eqnarray}
and
\begin{eqnarray}
I_4=(\frac{2k_g c_s^2}{S^2}+\frac{4k_g u_A^2}{3S^6})^2
\end{eqnarray}
The right hand side of Eq.(45) is smaller than the particle
(Peacock J A 2000) horizon of comoving radius $r_{H}=c\int
\frac{dt}{S}$. Hence, in a small fraction of the universe radius,
there exists an interaction between the GW and the plasma
producing small fluctuations with amplitude $P=h_{10}\exp{i[n_i
t-\Delta k z]}\frac{H_0^2}{S^3}[\frac{B_r+i B_i}{A_r+i A_i}]$.
During the interaction, energy of the order $P^2 L^3$ is
transferred from the GW to the fluid and magnetosonic waves are
produced. These magnetosonic waves do not get damped since in our
model we do not include dissipating processes. In a future paper
we intend to study a more realistic model and investigate further
astrophysical implications.
\newpage
{\bf 3.Discussion}

In the present paper, we have applied the equations governing finite amplitude  wave propagation
in hydromagnetic media discussed by Papadopoulos and Esposito (Papadopoulos D Esposito F P 1982),
in the so called Cowling approximation
in a perturbed spatial flat FRW cosmological model. We have assumed that, despite the presence of the magnetic field and
the perturbation $h_1$, the homogeneity and isotropy of the considered cosmological model do not change.
Upon the consideration of those equations, we have derived a dispersion relation (Eq.(29)) in the
radiation era (plasma epoch) .
We found that, if the gravitational waves and the magnetic field (produced after inflation period) are parallel,
then the gravitational waves may excite fluctuations which appear as fast magnetosonic waves.
From Eq.(29), it is evident that both, the magnetic field and the gravitational wave should
be different from zero otherwise we do not have genaration of density fluctuations.
The amplitude of the magnetosonic wave is proportional to the gravitational wave and does
not exhibit a damping since in
our model we do not take into account dissipative processes. The process we suggest seems to be reversible
in the sense
that energy is transfered from the gravitational wave to the plasma exciting magnetosonic waves  and vise versa.
The transfer of this energy from one side to the other occurs without loses and is of the order $P^2 L^3$.
Also, we have found that in an expanding universe, in the absence of the gravitational wave, a new
imaginary part of the frequency $(n_i)$ appears and the
real part is shifted by a term proportional to $\theta^2$. In the presence of the gravitational wave, the interaction of the plasma
with the gravitational wave produces magnetosonic waves oscillating at a frequency $n_r=n_g=k_g c$.
Finally, the application of the above equations to the study of small disturbances is rather straightforward although
tedious at this time,
but we hope to be able to discuss, in a future publication, these equations in more realistic hydromagnetic media.

{\bf Acknowledgements:} The author would like to thank Prof.
Loukas Vlahos, Dr. Kosta Dimopoulos and the Lecturer Nik
Stergioulas for there comments, criticism and beneficial
discussions. Also, the author would like to thank the referees
for their constructive comments and suggestions.

\end{document}